\documentclass[sigconf,screen,nonacm]{acmart}

\usepackage{xspace}

\definecolor{firstSink}{HTML}{1F77B4}
\definecolor{secondSink}{HTML}{2CA02C}

\newcommand{\firstSink}{{\textit{\textbf{\textcolor{firstSink}{first sink}}}}\xspace}
\newcommand{\secondSink}{{\textit{\textbf{\textcolor{secondSink}{second sink}}}}\xspace}

\AtBeginDocument{%
  \providecommand\BibTeX{{%
    \normalfont B\kern-0.5em{\scshape i\kern-0.25em b}\kern-0.8em\TeX}}}

\setcopyright{acmcopyright}
\copyrightyear{2018}
\acmYear{2018}
\acmDOI{XXXXXXX.XXXXXXX}

\acmConference[Conference acronym 'XX]{Make sure to enter the correct
  conference title from your rights confirmation emai}{June 03--05,
  2018}{Woodstock, NY}
\acmBooktitle{Woodstock '18: ACM Symposium on Neural Gaze Detection,
 June 03--05, 2018, Woodstock, NY} 
\acmPrice{15.00}
\acmISBN{978-1-4503-XXXX-X/18/06}

\begin{document}

\title{Market Interventions in a Large-Scale Virtual Economy}


\author{Senan Hogan-Hennessy}
\email{seh325@cornell.edu}
\affiliation{%
  \institution{Cornell University}
  \city{Ithaca}
  \state{New York}
  \country{USA}
}

\author{Peter Xenopoulos}
\email{xenopoulos@nyu.edu}
\affiliation{%
  \institution{New York University}
  \city{New York}
  \state{New York}
  \country{USA}
}

\author{Claudio Silva}
\email{csilva@nyu.edu}
\affiliation{%
  \institution{New York University}
  \city{New York}
  \state{New York}
  \country{USA}
}

\renewcommand{\shortauthors}{Hogan-Hennessy, Xenopoulos, and Silva}

\begin{abstract}
  Massively multiplayer online role-playing games often contain sophisticated in-game economies.
Many important real-world economic phenomena, such as inflation, economic growth, and business cycles, are also present in these virtual economies.
One major difference between real-world and virtual economies is the ease and frequency by which a policymaker, in this case, a game developer, can introduce economic shocks.
These economic shocks, typically implemented with game updates or signaled through community channels, provide fertile ground to study the effects of economic interventions on markets.
In this work, we study the effect of in-game economic market interventions, namely, a transaction tax and an item sink, in Old School RuneScape.
Using causal inference methods, we find that the tax did not meaningfully affect the trading volume of items at the tax boundaries and that the item sink contributed to the inflation of luxury good prices, without reducing trade volume. 
Furthermore, we find evidence that the illicit gold trading market was relatively unaffected by the implemented market interventions.
Our findings yield useful insights not only into the effect of market interventions in virtual economies but also for real-world markets.

\end{abstract}




\maketitle

\section{Introduction}
\label{sec:introduction}
Massively multiplayer online role-playing games (MMORPGs) attract hundreds of thousands of daily players.
In MMORPGs, players control a virtual character, and are able to move around a vast, open world with different activities and in-game items.
Like other online games, social interaction and skill advancement are central tenets in MMORPGs~\cite{castronova2001virtual}.
One crucial difference between MMORPGs and other online game genres is that each MMORPG typically maintains a robust in-game economy~\cite{lehdonvirta2005virtual}.
In these economies, there exist thousands of items, many of which are exchanged for in-game currency.
Thus, the many goods are dynamically priced according to the laws of supply and demand, rather than explicitly set by game administrators.
Therefore, virtual economies display some similarities to real-world economies~\cite{castronova2001virtual}.

Like real-world economies, MMORPGs witness economic phenomena such as inflation, illicit activity, and price manipulation~\cite{castronova2001virtual}.
In MMORPGs, these phenomena are especially important due to their direct impacts on player experience.
Game developers play a large role in managing in-game economies by implementing policy changes, in our case a market intervention.
For example, consider an item traded freely on the open market in a game.
If a game developer institutes a game update where a player is now available to buy the item from a non-player character (NPC) shop, then they have, in effect, instituted a price floor for the aforementioned item.
Game developers can also influence game economies by \textit{signaling} updates through community-focused communications, as well.
This is similar to how a central bank may signal interest rate changes~\cite{johnson2002effect}.
Increasingly, MMORPGs such as Eve Online~\cite{hillis_2007} and Old School RuneScape~\cite{jagex} have turned to economists to assess and manage their economies in recent years.

Inflation is an important economic phenomenon, with increasing relevance to both in-game and real-world scenarios.
In MMORPGs, it is common to see both inflation or deflation~\cite{castronova2001virtual,DBLP:journals/nms/CastronovaWSRXHK09,castronova2008synthetic,jung2011analysis}.
The former is often times caused by a lack of ``gold sinks'', which are in-game mechanisms to remove currency from circulation.
These sinks can take various forms, such as through taxes or fees.
Inflation may be further exacerbated by increasing numbers of high-level players who produce large amounts of gold or by illicit currency sellers who use massive bot farms to produce gold.
On the other hand, deflation can become a problem when a game does not provide adequate currency-generating activities~\cite{wright_2021, colbert_2021}.
In such an instance, players may resort to bartering as opposed to using currency as an exchange mechanism.
Nonetheless, both inflationary and deflationary environments have the potential to erode player experience, and are thus crucial to manage~\cite{harambam2011game}.

MMORPG economies offer a unique window to study economic policy due to their frequent updates and markets that resemble the real-world~\cite{DBLP:journals/gamestudies/Castronova03}.
In this paper, we explore the results of a natural experiment in Old School RuneScape (OSRS).
Specifically, we analyze the effects of two economic interventions, a transaction tax and an item sink, in OSRS.
We employ various quasi-experimental approaches, such as difference-in-difference, regression-discontinuity and regression-kink designs.
Furthermore, we discuss the implications of the tax on real-world trading.
We find that the item sink did increase the prices of the targeted items. 
On the other hand, trading volume was relatively unaffected by both the transaction tax and item sink. 
Furthermore, we observe that real-world trading prices were relatively unaffected by the market interventions.
Our findings are not only useful to understanding and managing virtual economies, but also may provide a basis for understanding real-world economic phenomena through natural experiments found in many virtual economies.

The rest of the paper is structured as follows.
In Section~\ref{sec:related-work}, we review the related literature regarding MMORPGs and virtual economies.
Section~\ref{sec:runescape} details RuneScape and its data.
Furthermore, we analyze the dynamics of the RuneScape economy.
In Section~\ref{sec:ge-tax}, we outline the natural experiment, our quasi-experimental method, and analyze the impacts of the transaction tax.
Section~\ref{sec:discussion} provides further discussion on our analysis, provides recommendations for natural experiments in MMORPG economies, and discusses limitations and future opportunities.
Furthermore, we outline how natural experiments in virtual economies may be useful for game developers.
Finally, we conclude the work in Section~\ref{sec:conclusion}.


\section{Related Work}
\label{sec:related-work}


\subsection{Virtual Economies}

Virtual economies, as a research domain, largely gained prominence due to MMORPGs in the early 2000s~\cite{castronova2001virtual}. There exist many clear analogues to real-world economic phenomena in virtual economies. Although there remain some differences, Castronova asserts that virtual economies may still provide a ripe environment for understanding real-world economies~\cite{DBLP:journals/gamestudies/Castronova03}. Many of the same quantities studied in economics may also be calculated and studied in virtual economies. For example, Castronova provides a macroeconomic summary of EverQuest~\cite{castronova2001virtual}. In their summary, they detail topics such as exchange rates between real-world and virtual currency, per-capita production, wages, inequality, and inflation. For \textit{EverQuest 2}, Castronova~et~al. observe that aggregate economic behavior follows closely with that of the real-world; however, fluctuations are more dramatic~\cite{DBLP:journals/nms/CastronovaWSRXHK09}. In particular, they find high rates of inflation in some virtual economies.

Inflation is a core economic concept that concerns the increase of the price of goods and services. In online games, there exists a similar concept of \textit{mudflation}~\cite{castronova2008synthetic}. Mudflation, an amalgamation of multi-user dungeon (MUD, a precursor genre to MMORPGs) and inflation, is a phenomenon in MMORPGs where prices of once expensive goods decline in value.
Harambam~et~al. note that concerns over mudflation are noted not only by researchers but also by players themselves~\cite{harambam2011game}.
Price swings have significant effects on player retention. For example, Jung~et~al. note that deflation may cause boredom among players~\cite{jung2011analysis}. On the other hand, inflation can significantly degrade player experience by making items unattainable, thus exacerbating inequality in a game. Inflation in a virtual game economy is largely determined by in-game economic mechanisms that regulate the production and consumption of currency~\cite{he2018study}. Lahti argues that if a player earns the most from trading, rather than generating new wealth, then inflation should stay low. They further suggest that the economy must be structured such that it utilizes low value items, regardless of player tenure~\cite{lahti2015inflation}. Lehdonvirta and Castronova suggest that a monopsony market structure, where the game publisher buys goods from players, addresses the issue of deflation~\cite{lehdonvirta2014virtual}. Furthermore, they note that trading allows players to play according to their competitive advantages, and thus allows a player to focus on aspects of the game they enjoy most. Virtual economy inflation may also be affected by characteristics of the player base. For example, Belaza~et~al. suggest that players from countries with high real-world inflation tend to buy more than sell in-game, thus mimicking real-world behavior~\cite{belaza2020connection}. 

Like policymakers in the real-world, game operators may introduce changes to the game economy through game updates. Also akin to the real-world, these policy changes may deliberately, or indirectly, morph the economy of a game. For example, Castronova highlights the case of \textit{Everquest}, where significant item deflation was observed in items after game expansions~\cite{castronova2001virtual}. For RuneScape, Bilir suggests that, in light of a game update which eased item trading costs, the overall price index decreased~\cite{bilir2009real}. Game publishers may also directly impose economic policy, such as through price floors or ceilings. Castronova proposes that price ceilings do not generate excess demand and that price floors do not create excess supply~\cite{DBLP:journals/gamestudies/Castronova03}. Castronova further suggests that it may useful to directly control the prices of some goods. In games like RuneScape, in-game mechanisms, such as skills, provide price floors for certain goods. Game developers may test the effects of different policies through A/B testing on dedicated servers~\cite{castronova2015policy}. Stephens~et~al. propose a reinforcement learning approach to simulate the game economy to test the effects of updates on inflation~\cite{DBLP:conf/icaart/StephensE21a}.

We build on prior work in the following directions.
First, we utilize the market intervention aspect of game updates in a virtual economy.
Specifically, we concern our analysis with the impact of a transaction tax and an item sink in Old School RuneScape, a popular MMORPG with tens of thousands of players.
Furthermore, we employ multiple econometric methods to parse the independent causal effects of multiple market interventions, including \textit{difference-in-differences}, \textit{regression-discontinuity} and \textit{-kink} designs.
Lastly, we assess the effectiveness of the policy changes on item prices and trade volume relative to the policy goals of the game publisher.

\subsection{Social Behavior in MMORPGs}
Many MMORPGs have a high degree of player interaction, such as through text chat or trading mechanisms. These interactions have important effects on player retention and economic activity. For example, Jeong~et~al. find that friendship and trading interaction networks have high degrees of overlap compared to other types of social interactions~\cite{DBLP:conf/sigcomm/JeongKK15}. In \textit{Aion}, a popular MMORPG, Chun~et~al. observe that interactions such as message exchange are highly correlated with trade~\cite{DBLP:conf/www/ChunCHK018}. Kang~et~al. show that people tend to join guilds, which are in-game groupings of players, when they expect benefits, such as money or valuable items~\cite{DBLP:conf/www/KangPLK15}. Furthermore, users often exhibit a profit-seeking mentality by switching guilds. Bisberg~et~al. find that players who experience or observe generosity in MMORPGs generally show higher future game engagement~\cite{bisberg-mmo}. Chung~et~al. observe that when a player group is not cohesive, there tends to be significant attrition~\cite{DBLP:conf/www/ChungHCKKC14}. On the other hand, players who gather large amounts of currency or valuable items are less likely to churn from the game~\cite{DBLP:conf/www/ParkCKC17}. 



\subsection{Real-World Trading}

Real-world trading (RWT) is an activity, often against most online games' terms of service, where a player trades real-world currency for an in-game currency, item or service. The trade between real-world currency and in-game currency has existed since MMORPGs were first popularized~\cite{castronova2001virtual}. Oftentimes, large bot farms are used to generate currency for RWT~\cite{DBLP:conf/ndss/LeeWKMK16}.
Wohn~et~al. find that large friend networks, along with high rates of virtual good exchange, were positively associated with a user engaging in RWT~\cite{DBLP:conf/chi/Wohn14}.
Furthermore, they find that users who engaged in high amounts of RWT were primarily buying items for their visual effects, whereas low spenders were buying consumable items for playing the game.
Similarly, Xu~et~al. find that server population and network density are correlated with in-game currency price~\cite{DBLP:journals/intr/XuYLLPWC17}.

The effect of RWT on MMORPGs is disputed.
Lehdonvirta suggests that some subsets of players may see RWT as beneficial, such as achievement and immersion-oriented players~\cite{lehdonvirta2005real, lehdonvirta2005virtual}. Likewise, Jung~et~al. suggests that RWT is beneficial to game operators and that there exists an optimal amount of supply for game items~\cite{jung2011analysis}.
On the other hand, Huhh suggests that RWT may have an impact on player retention, and points to the demise of \textit{Lineague 2} as an example of RWT's effect on player experience~\cite{huhh2005empirical}.

The size of RWT in a game can be large. For example, Chun~et~al. show that over half of trades involving upper-class Aion players are associated with RWT entities~\cite{DBLP:conf/www/ChunCHK018}. Due to the perceived harm that botting, the activity of using an automated program to play the game, may have on the game, many approaches have been adopted to identify bots and accounts involved in RWT. Fujita~et~al. exploit the trading network of a game to calculate the likelihood of users being involved in RWT activity~\cite{DBLP:conf/aiide/FujitaIM11}. Similarly, Jeong~et~al. find that social interaction networks varied substantially between bots and human players~\cite{DBLP:conf/sigcomm/JeongKK15}. Lee~et~al. construct a trading network and decompose it into various communities~\cite{DBLP:conf/www/LeeWKK18}. They find that professional RMT providers typically form specific network structures, such as star-shaped or chain networks. Finally, Tao~et~al. use machine learning to identify RWT associated accounts~\cite{DBLP:conf/kdd/TaoLZZWFC19}.

\section{RuneScape Markets}
\label{sec:runescape}
Old School RuneScape (OSRS) is the latest rendition of the long-running RuneScape MMORPG series, which was first released in 2001.
Functionally, OSRS retains many characteristics of the 2007-era version of RuneScape, remaining visually and mechanically similar, but has received regular updates.
In this section we describe OSRS, its in-game market, and its data.

\subsection{Game Description}
OSRS, which typically has at least 50,000 accounts logged in at any given time, is an MMORPG set in a large and open fantasy world.
In this world, players can gain experience towards a variety of in-game skills and interact with other players.
Skills are generally broken down into four categories: (1) combat skills, (2) gathering skills, (3) artisan skills, and (4) support skills.
While progressing most skills requires some form of item consumption, skills may also produce many items as a byproduct of their progression. 
These items are often exchanged for gold pieces (``GP'') with in-game shops, through NPCs, or with other players.
For example, training one's ``woodcutting'' skill produces logs. 
These logs may be purchased by players who wish to train their ``fletching'' or ``firemaking'' skills.
Thus, there is a significant amount of player interaction in OSRS, especially regarding the exchange of items. 
Although OSRS is free-to-play, many players opt to buy monthly membership, which allows for much broader content, including new areas, items, and skills.

\subsection{The Grand Exchange}
\begin{figure}
    \centering
    \includegraphics[width=\linewidth]{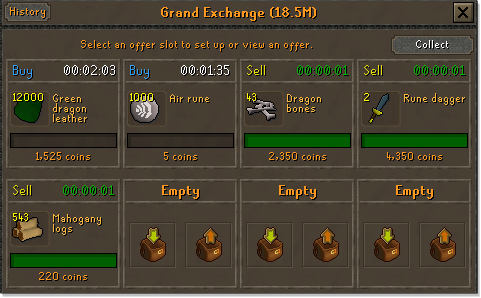}
    \caption{The Grand Exchange (GE) is a mechanism within RuneScape to connect buy and sell orders for thousands of items.}
    \label{fig:grand-exchange}
\end{figure}

The Grand Exchange (``GE'') is a marketplace that connects buyers and sellers of all OSRS tradeable items, similar to a real-world securities exchange.
The GE works by clearing offers between buyers and sellers in the order it receives them.
For example, suppose a buyer places an order for an item at $P_{B}$ and there exists a sell offer at price $P_{S} < P_{B}$.
Then, the buyer receives the item at price $P_{S}$ and also receive $P_{B} - P_{S}$ in gold; the seller receives $P_{S}$.
If a seller places an order at price $P_{S}$ and a buy offer exists at price $P_{B} > P_{S}$, then the seller receives $P_{B}$ gold in exchange for the item.
Thus item prices are determined by the market, and vary over time.
Items also have a buy limit, which restricts the quantity of an item that a player can purchase within a four hour game window.
There is no sell limit.
We show an example of the GE interface in Figure~\ref{fig:grand-exchange}.
In total, a player who pays for a RuneScape membership may have up to eight concurrent buy and sell orders.

\subsection{Dataset}
OSRS contains roughly four thousand items that are able to be traded on the GE.
Jagex, OSRS's publisher, provides prices for these items on the official game website, but these prices are generally delayed, sometimes by as much as 24 hours.
Conversely, the OSRS Wiki, the official community wiki, provides an endpoint which returns the average instant-buy and instant-sell prices and volumes for a given item.
To do so, they leverage a partnership with RuneLite, a popular, open-source third-party client used to play OSRS.
RuneLite is officially recognized by Jagex as one of three approved third-party clients.
At the time of writing, roughly 75\% of the player population used RuneLite.
Whenever a player using RuneLite bought or sold an item on the GE, we collected the average daily buy and sell prices and trade volumes for all items from August 2021 to August 2022.


\subsection{RuneScape Economy}
\label{sec:price-graph}
As noted by Lehdonvirta and Castronova, players oftentimes play towards their own competitive advantages, and thus are likely unable to produce all items that meet their consumption needs~\cite{lehdonvirta2014virtual}. 
Thus, the OSRS economy has notable similarities to real-world economies: players work towards producing items or achieving goals and they trade for products they want or need.
The GE is the primary mechanism where players sell what they produce, and trading on the GE represents the majority of economic activity in OSRS~\cite{bilir2009real}.

Our data cover roughly four thousand items in the OSRS universe; however, only about 70\% of these items are traded on a daily basis. 
Concerning this subset of items, we calculate the correlation in price movements between all items.
Specifically, we find that, on average, an item has a correlation $|\rho| > 0.5$ with roughly 200 items.
When we change the cutoff to $|\rho| > 0.2$, the number of correlated items rises to around 1,100.
Thus, we find that many items exhibit strong price correlation with other items. 
This is expected, as many items are purchased jointly. 
For example, consider the item ``Dragon Crossbow,'' a popular in-game weapon.
The Dragon Crossbow has a correlation of $0.86$ with the item ``Armadyl d'hide body,'' an armor piece that gives a bonus to using the Dragon Crossbow.
Thus, the two goods are complementary, and therefore one would expect a high degree of (positive) price correlation between the two.

In aggregate, our data indicate an average of 4.5 trillion GP worth of items exchanged over the exchange, daily.
The bond, the sole in-game item that a player may purchase with real-world currency, can also be bought and sold on the Grand Exchange for GP.
Bonds may be used in-game to purchase ``membership,'' which unlocks additional content.
If we assume an average GP price of \$1.45, which is roughly the price as determined by a bond, then the implied daily volume of trade is \$6.5 million.
The implied yearly volume is thus \$2.3 billion, which is larger than the gross domestic product of some countries.
Furthermore, it is also important to note that our data are only reflective of the orders made by 75\% of accounts, and thus our estimate is likely an underestimate of the true daily transaction volume in OSRS.

\begin{figure*}[h!]
    \includegraphics{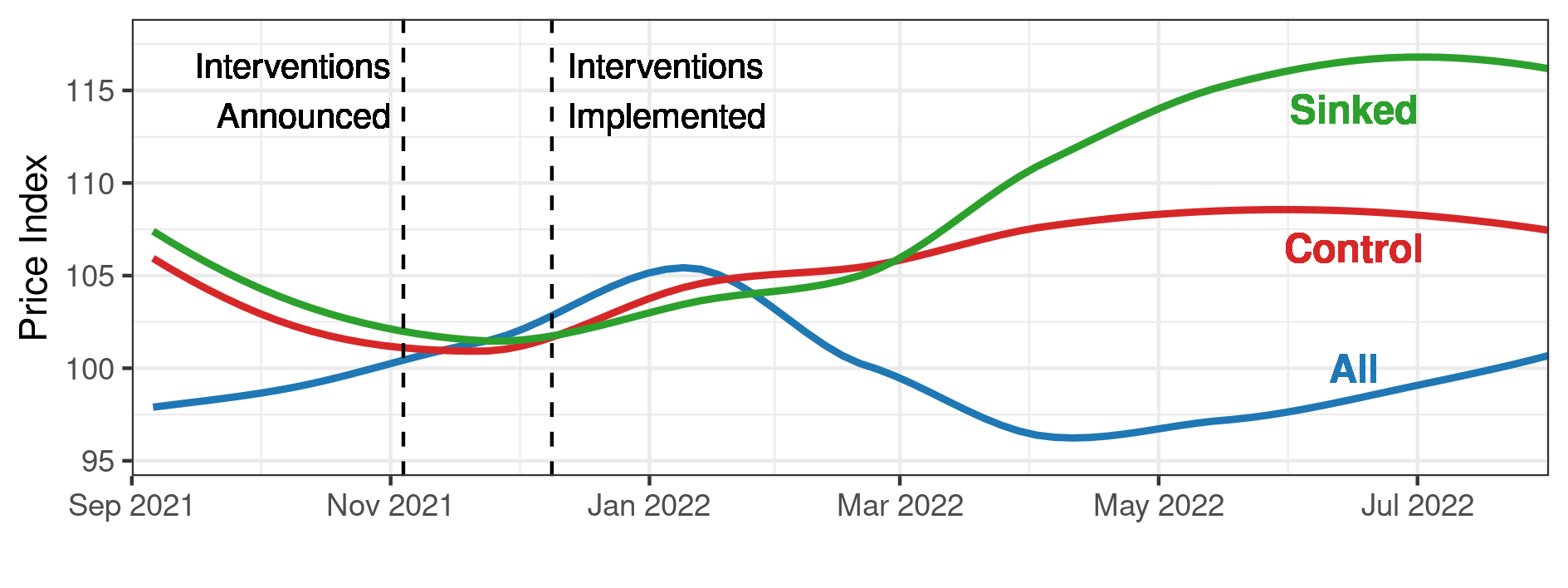}
    \caption{Item Price Index across all items traded on the OSRS Grand Exchange.
        A level of 100 indicates the prices at 08 December 2021.
        ``Sinked'' refers to the set of items affected by the 09 December 2021 market intervention (sink) and ``Control'' refers to a set of similar, highly priced items that are unaffected by the sink nor are highly correlated with sinked items.}
    \label{fig:price-index}
\end{figure*}

We can measure classic macroeconomic phenomena using the OSRS GE data.
A \textit{price index} tracks the price of a basket of goods, relative to a point in time.
In the real-world, governments calculate price indices to track price inflation.
Using our GE data, we construct a price index by calculating the mean price of goods (weighted by trade volume) in each week, and compare each week's percent change in price relative to the week preceding new market interventions.
To construct the price index, consider $P_{k,t}$ for an item $k$'s sale price in week $t$, $w_{k,t}$ its trading volume, and $t_0$ the base week to compare to.
Then, we construct the price index $\widetilde P_{K, t}$ for a group of items $K$ in week $t$ by comparing the weighted average prices $\overline P_{K, t}$, as in equation~\eqref{eqn:priceindex}.
In practice, we perform this calculation via linear regression, separately for three item groups, as shown in \autoref{fig:price-index}.

\begin{equation}
    \label{eqn:priceindex}
    \overline P_{K, t} = \frac{\sum_{k \in K} P_{k,t} w_{k,t}}{\sum_{k \in K} w_{k,t}}, \;\;
    \widetilde P_{K, t} = 100 \left( \frac{\overline P_{K, t} }{ \overline P_{K, t_0} } \right)
\end{equation}

Item prices were, for all items, rising for the final months of 2021, yet the first few months of 2022 saw the average price fall of 10 percentage points, and relatively stable prices in the middle of 2022, changing no more than 5\% in aggregate from April 2022 to August 2022.
On the other hand, items which were affected by a new trading sink, which we describe in depth in Section~\ref{sec:ge-sink}, experienced falling prices prior to the market intervention.
Afterwards, these items stabilized and gradually increased post-intervention.
In Section~\ref{sec:ge-interventions}, we discuss and analyze the market interventions in detail.

\section{Analyzing Market Interventions}
\label{sec:ge-interventions}
Jagex announced two large market interventions to trading on the GE in November 2021: a transaction tax and an item sink~\cite{jagex}.
Both market interventions were formally introduced in December 2021, and have stayed active to date.
Jagex had publicly expressed concern that high-level items, items limited to players with a lot of game experience (and GP), were too common-place.
And so the company implemented a two-pronged market intervention to tackle the issue of item pricing. 
First, the transaction tax collects 1\% of the item sell price (subject to restrictions) for every GE transaction, thus impacting the profitability of real-world traders and botters who use the GE at high volume to generate GP. 
The tax is taken out of the seller's proceeds. 
Second, the proposed item sink uses the proceeds from the transaction tax to remove high-level items by directly buying the items from players on the GE. 
If the virtual coffers filled by the transaction tax are insufficient, the item is not removed.
The number of items to be removed and deleted is subject to a daily maximum set by Jagex. In this section, we use quasi-experimental techniques to analyze the effects of the tax on item prices and trading volumes.
We do so by exploiting discontinuities in the tax rates, comparing sinked items to a plausible control group, and inferring trends in real-world trading via the price of illict in-game currency purchases.

The combined policy change is largely similar to real-world market interventions.
A financial transaction tax is a popular economic policy that regulators employ to discourage market volatility and speculation, by increasing the price of an additional trade, and thus reducing trade volume.
Many countries levy a financial transaction tax in their securities exchanges, and the European Union recently considered a Union-wide financial transaction tax~\cite{rodriguez2021financial}.
The proposed item sink mirrors government efforts to remove items from circulation via market mechanisms, such as the Australian government's fire-arms buyback in 1996~\cite{bnrew_2017}.
Lastly, the tax and sink are redistributive in nature, since proceeds of the transaction tax are then passed back to the public (in this case, the player base) via another program.
This is in line with the economic policy approach of tying a tax to another policy provision~\cite{golladay2013economic}, such as ear-marking proceeds from gambling taxes to fund education~\cite{simmons2006gambling}.

\subsection{Transaction Tax Effects}
\label{sec:ge-tax}
\begin{figure}[h!]
    \includegraphics[width=\linewidth]{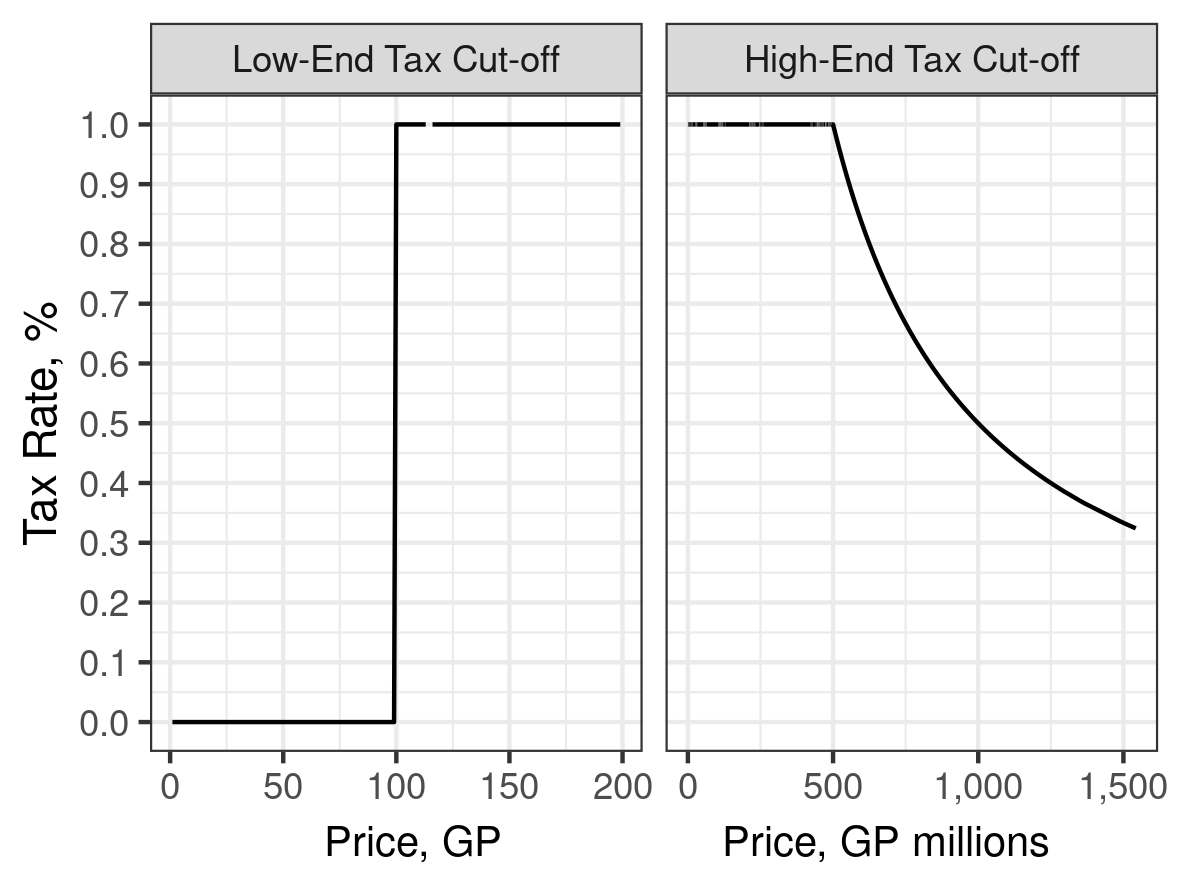}
    \caption{Tax rate as a function of an item's sale price.
        Items under 100 GP suffer no tax, and max nominal tax is capped at 5 million GP.}
    \label{fig:taxrate}
\end{figure}

We analyze the causal effect of the transaction tax on item price and trade volume by exploiting the non-linear jumps in the tax rate.
The tax is 1\% of the item sale price, but only applies if the price is greater than 100 GP. The tax is capped at 5 million GP regardless of item sell price. 
This leads to two distinct discontinuities: (1) a jump in the tax rate of sales above and below 100 GP, and (2) a change in the slope of the tax rate above and below 500 million GP.
These first-stage discontinuities are exhibited in \autoref{fig:taxrate}.
We use this approach to \textit{identify} the effect of the transaction tax; the method produces point estimates for the causal effect at the tax cut-off points \cite{hahn2001identification}.
Notably, provision of the tax is decided by an item's price, so we cannot disentangle the causal effect of the tax on item price, and thus we focus on the effect on trading volume.
Alternative methods, such as a correlation analysis, would only be able to show the association between the policy change and economic outcomes, but our approach allows for the analysis of the causal relationship.
Unless otherwise noted, our analysis considers the outcomes with a log transformation.
The log transformation is monotonic, so does not meaningfully affect results, and allows for a percent-change interpretation of the estimated effects.

There is a sharp-discontinuity in tax rate at 100 GP, so we estimate the average effect of the 1\% transaction (on items with sale price of 100 GP) through a regression discontinuity (RD) design~\cite{hahn2001identification}.
This approach assumes that items are virtually the same, apart from the tax applied at the cut-off, in a neighborhood around the 100 GP cut-off.
Around 100 GP, the effect of the transaction tax on trading volume is not distinguishable from zero at any standard level of significance.
\autoref{fig:after-tax} shows the graphical representation of the results, where the estimated discontinuity in volume above and below 100 GP is not distinguishable from zero.


Secondly, there is a discontinuity in the rate of change of the tax rate at 500 million GP, so we estimate the average effect of the 1\% transaction (on items with sale price of 500 million GP) through a regression kink (RK) design~\cite{card2017regression}.
This approach again assumes that items are virtually the same apart from a differing tax rate, in a neighborhood around the 500 million GP cut-off.
Around 500 million GP, the transaction tax of 1\% reduced trading volume by 6.9\%, yet bounds on the treatment effect range -13 to -1\% at a 95\% significance level. 
This means that, while the reduction on trading volume is accepted as non-zero, our exact estimate is not precise, and could reasonably be as large as -13\%, or as small as -1\%.

\subsection{Item Sink Effects}
\label{sec:ge-sink}

We analyze the causal effect of the new item sink on item price and trading volume by comparing the items that the new sink applied to against a plausible control group of items.
The difference-in-difference (DiD) method~\cite{card1994,bertrand2004much}, popular in empirical economics, generalizes the insight of John Snow's 1856 cholera study~\cite{snow1856mode} by comparing the \textit{difference} in treatment effect between treatment and control groups before and after a new policy.
DiD allows for a causal comparison in outcomes between a treatment group and a control group when researcher-directed randomization did not occur, as is the case of the item sink.
In our case, the treatment group is the items affected by the sink, and we address concerns for treatment selection by constructing a control group of items which the sink could have reasonably applied to.
Section~\ref{sec:did} provides further technical details of the DiD method.

Jagex implemented the item sink in two rounds. 
The first round on 09 December 2021 started buying and removing 34 items, all of which are high-level items which only very experienced players can afford and use. 
The second round started additionally buying 27 more high-level items on 16 December 2021.
We treat each round as a separate policy event by comparing the items in the first round and second round separately to a control group that we construct. 
Given a set of all high-level items $\mathbf{I}$, all of which have an average price greater than $100,000$ GP, and the set of sink-targeted items $\mathbf{I}_{\text{sink}} \subset \mathbf{I}$, we define the control set
\begin{equation}
    \label{eqn:control}
    \mathbf{I}_{\text{control}} = \left\{
        i \in \mathbf{I} \setminus \mathbf{I}_{\text{sink}} \mid 
            |\rho(i, k)| < 0.1, \; \forall k \in \mathbf{I}_{\text{sink}} 
                \right\}
\end{equation}
\noindent where $\rho(i,k)$ measures the price correlation between items $i$ and $k$.
The construction infers which items show no significant price correlation (positive or negative) with the sinked items, and thus are neither reasonable substitutes nor complements to the sinked items.
By construction, this control set represents similarly high-priced items to those in the sink, yet changes in demand motivated by the new item sink should not otherwise affect demand for items in the control group.

\begin{figure}[h!]
    \includegraphics[width=\linewidth]{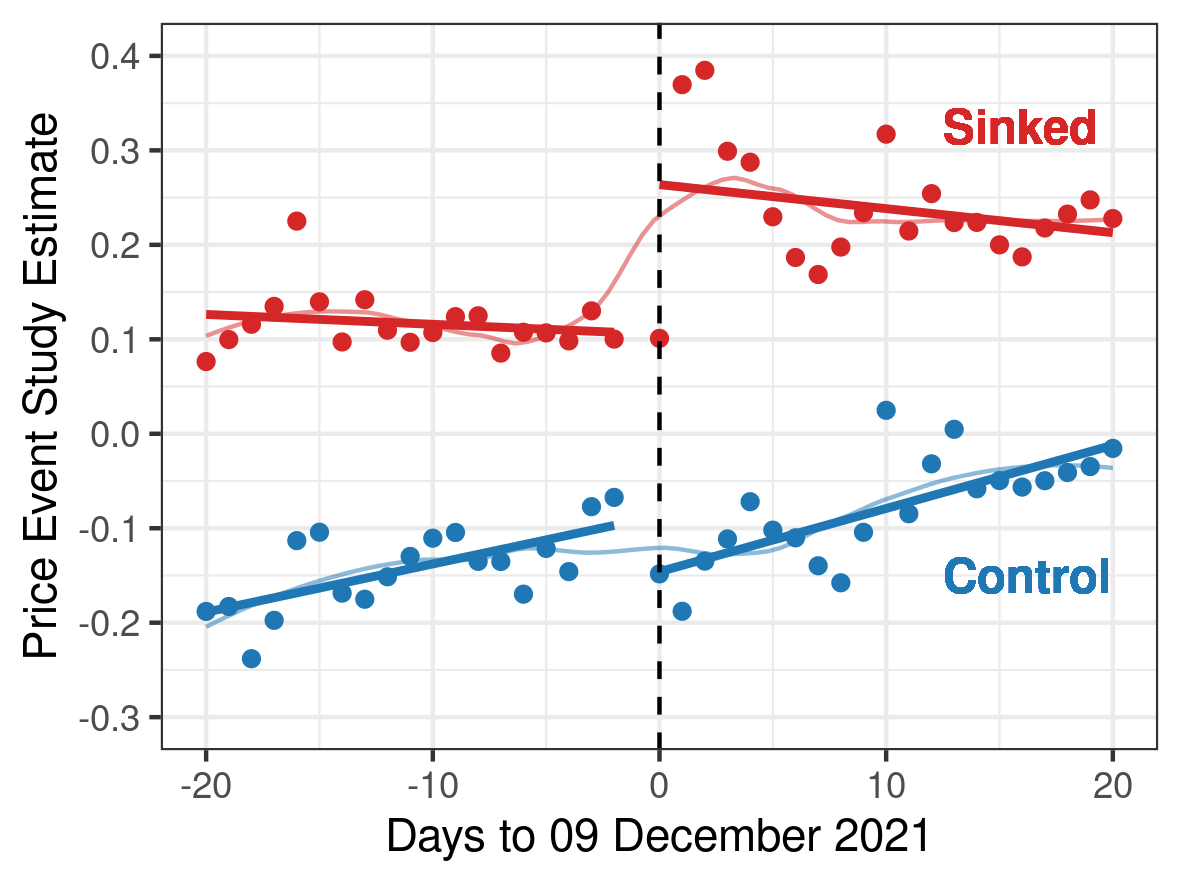}
    \caption{Pre-Trends in item price for $\mathbf{I}_{\text{sink}}$ and $\mathbf{I}_{\text{control}}$, around the 09 December 2021 intervention date.}
    \label{fig:price-pretrends-sink}
\end{figure}

To verify our approach, we investigate the time trend in price before and after the market intervention, for both $\mathbf{I}_{\text{sink}}$ and $\mathbf{I}_{\text{control}}$ in \autoref{fig:price-pretrends-sink}. 
Prior to the first sink on 09 December 2021, the sinked and control items do not have trends distinguishable from each other, and the control group is not meaningfully affected by the new item sink.
Although the trend for control group prices appear positive, the trend is small in magnitude and thus the associated standard errors of the point estimates consistently contain zero, and the trend is not statistically distinguishable from no-trend.
The lack of distinguishable trend between the groups validates the assumption that pre-trends are similar, thus justifying the DiD approach we employ to identify the causal effect of the sink on item prices.
By the same argument, we validate the pre-trends assumption for trading volume in \autoref{fig:volume-pretrends-sink}.

\begin{figure}[h!]
    \centering
    \includegraphics[width=\linewidth]{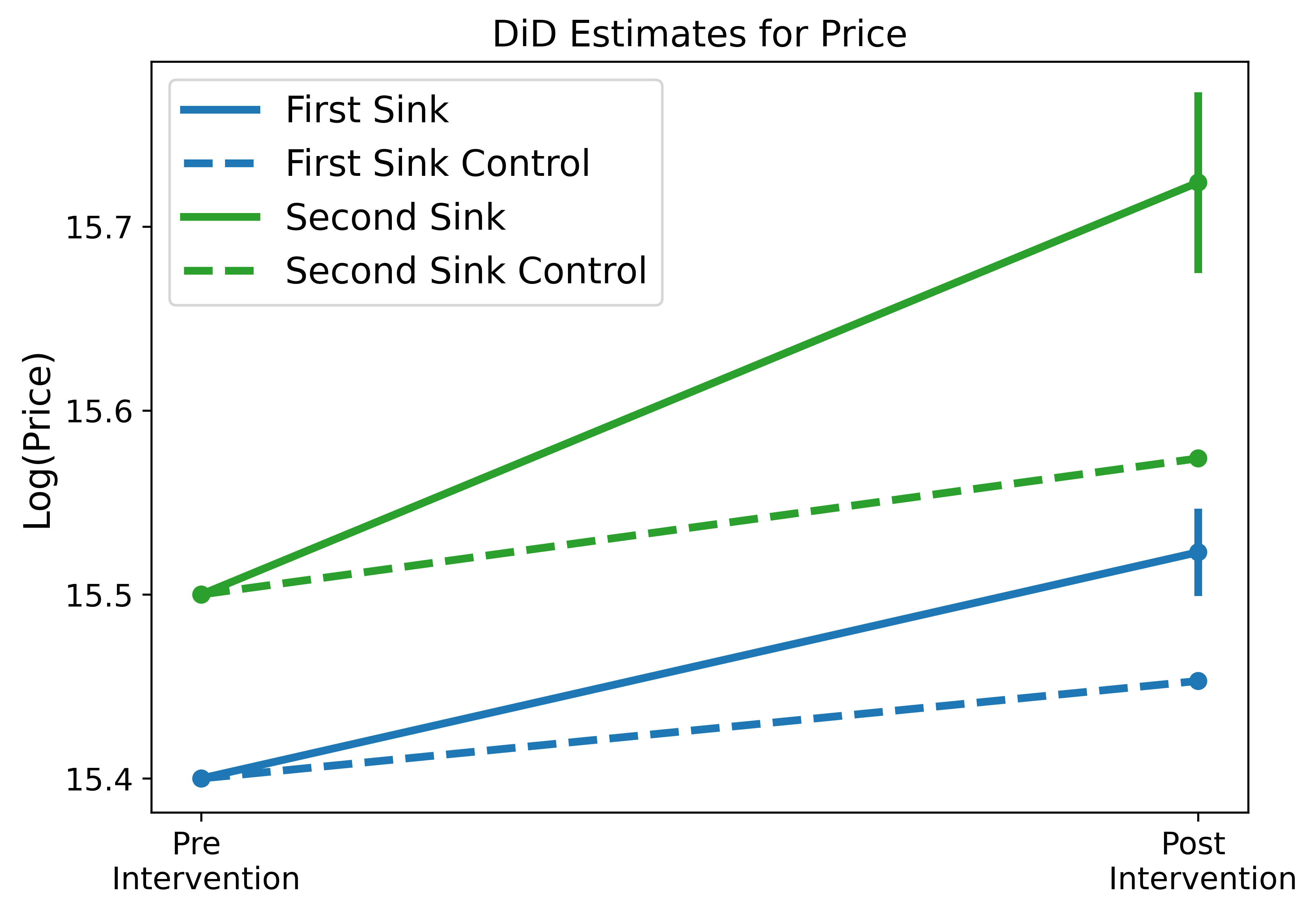}
    \caption{Lines are colored by \firstSink and \secondSink. Dotted lines indicate the counterfactual trend for market intervention, and vertical bars denote the confidence interval for the estimated effect. The item sink rose the price, but not volume, for sinked items.}
    \label{fig:did}
\end{figure}

The first treatment group experiences a jump in prices on the order of 7\% when the sink was implemented, with a confidence interval of 5.0\% to 9.6\% at the 95\% confidence level.
This corresponds to the jump in prices for the sinked group on 09 December 2021, minus the change for the control items, in \autoref{fig:price-pretrends-sink} (and we present full point estimates in the appendix).
We also estimate that items in the second sink experienced a jump in prices on order of 14.4\%, with a confidence interval of 11.9\% to 16.9\% at the 95\% confidence level.
On the other hand, the new item sink's effect on items' trading volume was indistinguishable from zero in both the initial and second round of the sink.
We show the visual representation of the DiD estimates for price in Figure~\ref{fig:did}, and for volume in Figure~\ref{fig:did-volume}.

\subsection{Real-World Trading}
\label{sec:rwt}
The last topic of analysis is real-world trading, where we infer interactions between real-world players and the virtual economy via different exchange rates for virtual currency.
Players can buy ``bonds'' from Jagex, and these bonds are in-game items that grant the player a 14-day game membership. 
Importantly, bonds may also be sold between players on the GE.
Thus, although the bond's price is fixed in the real-world (\$6.99 for our analysis period), the in-game GP price of the bond floats on the GE via an auction mechanism.
This gives us a measure of the varying price (in US dollars) for in-game currency (GP), from which we can infer if there have been any changes in demand from players.
Notably, buying bonds and selling them in-game for GP is explicitly allowed by Jagex's terms of service and does not fall under the realm of real-world trading.

\begin{figure*}[h!]
    \includegraphics{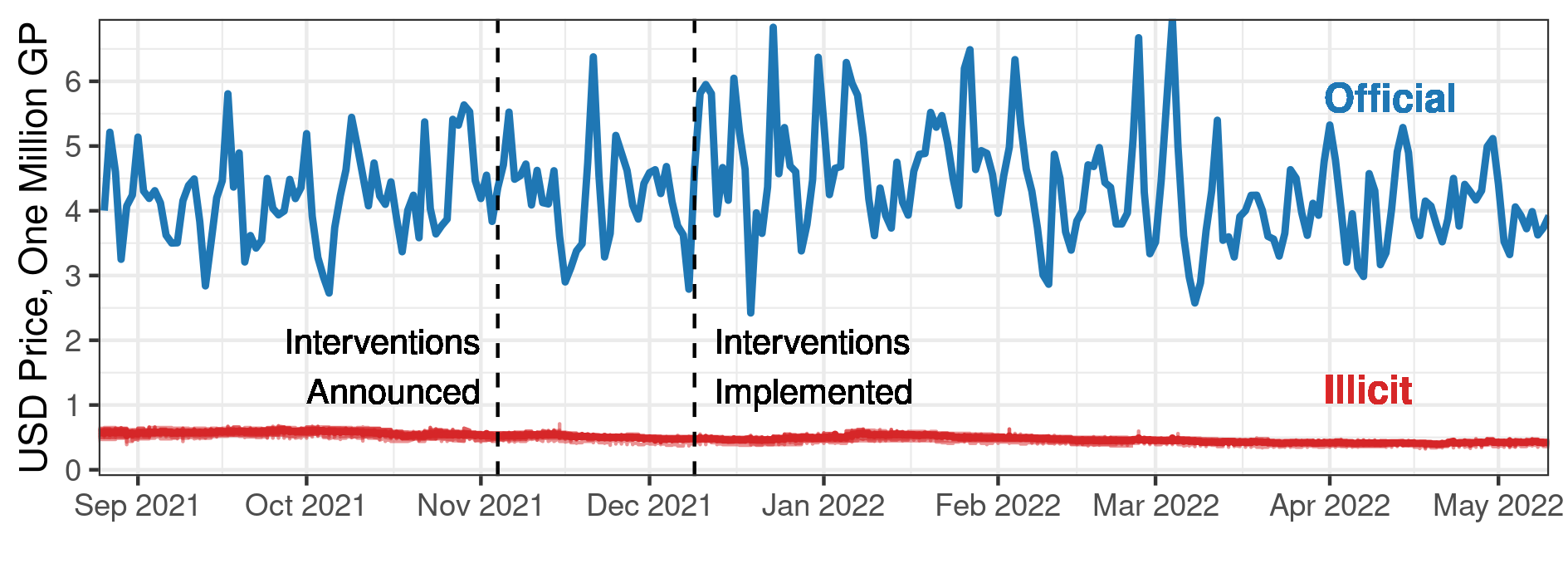}
    \caption{Official price versus the illicit market average price in \$-per-1m GP.
        Individual illicit GP sellers denoted by transparent lines, and in practice underlay the mean illicit price.
        Illicit GP prices remained stable after introduction of the market interventions, and official GP prices did not become more variable.}
    \label{fig:gp-price}
\end{figure*}

On the other hand, players can buy in-game GP from real-world traders \textit{illicitly} (i.e., breaking the OSRS terms of service).
Real-world traders exploit a number of methods to generate in-game currency in a cost-effective manner, from utilizing sophisticated procedures to automate in-game production (i.e. ``botting'') to employing operators to manually play different accounts.
Real-world traders typically generate huge sums of in-game GP by producing large numbers items, and then selling these items to other players on the GE for large sums of GP.
In doing so, real-world traders generate large numbers of items in a manner that is explicitly banned by Jagex. 
In relation to the items which are affected by the item sink, real-world traders increase the supply of items that Jagex has designed to be in limited supply, and thus decrease the market price for items that Jagex has designed to be very expensive.

These activities are illicit as they are explicitly banned by the game operator, Jagex, who has has been tackling real-world traders and botters for at least the last decade~\cite{guest2010jagex}.
Jagex is active in enforcing their terms and services and they often ban accounts for which they find evidence of botting or real-world trading activity, confiscating any items generated in the process.
Real-world traders sell their produced GP to players via illicit external websites; this is an illicit black-market, where scams and risk of a game ban is high.
Because of these market frictions, when compared to the official bond market, there is a risk premium for illicit GP purchases, measured by the difference in \textit{illicit} and \textit{official} price for GP (in USD).
As such, we collected the US dollar price-per-million GP from seven popular illicit gold sellers over our sample period, and compared their mean price to that in the official bond market. 

We see that the official price-per-million GP varies between \$2.50 and \$6.50, and exhibits no explicit long-run increasing or decreasing trend.
The official price is centered around a mean of \$4.17.
The illicit market has virtually no variation to speak of: a million GP sells on the black market at an average of around \$0.495 for our entire sample period, and all sellers' prices are between \$0.45 and \$0.58 -- a very tight spread compared to the official channel.
While the ten illicit sellers' prices varied between 40 and 60\textcent for a million GP across the entire time period, prices on both sides of the new market interventions reach both ends of the range before \textit{and after} the market interventions.
We tested for structural breaks in both time series, and found no statistically significant breaks in mean or variance.

Notably, over this time period, supply of GP in the official bond market saw no explicit changes.
Jagex did not change any policies for bond buying between September 2021 and May 2022, and most variation in the official price will have arisen from changes in player demand.
We do see short-run increases in the US dollar price for GP around both the market intervention announcements and the implementation.
But, the short-run increases are not particularly large, and are overshadowed by larger short-run changes in different months (where there was no other market intervention).

Although not a stated target of the market interventions, real-world traders and botters are affected by the market interventions due to an increased cost of operation through the GE transaction tax.
Every time a real-world trader sells items on the GE, they pay a transaction tax of 1\%, and thus face higher costs compared to the time before the transaction tax existed.
And yet, illicit real-world traders did not change the price they charged for in-game GP around the market interventions.
As such, we infer that real-world trading, economic interactions between the real-world and the in-game economy of OSRS, were not meaningfully affected by the market interventions.
\section{Discussion}
\label{sec:discussion}

In this study we applied methods from empirical economics to estimate the causal effects of new market interventions in a large-scale virtual economy.
We found that the the interventions did not uniformly affect economic outcomes: prices for high-level items increased due to the item sink, but trading volume was not meaningfully affected by the transaction tax.
Our results align clearly with standard economic theory: the new item sink increased demand for high-level items (by buying items), thus leading to increased prices.
And yet, gauging whether the market interventions were successful depends on the policy objective of the game planner.

\textit{Policy Objectives}.
Jagex has been monitoring and stabilizing its in-game economy since the early 2000s, and has been combating real-world trading since before then.
The 09 December 2021 market interventions were a very consequential change to the OSRS economy, but Jagex releases other game updates and additions regularly.
While it is clear that the transaction tax meaningfully reduces the amount of in-game currency in circulation, and the item sink reduces the number of (targeted) high-level items in circulation, Jagex has not publicly stated their intentions for this round of market interventions.
Without a definite policy target, it is not possible to state whether the game designer achieved their policy goal with these market interventions.
Thus, we leave it for future work to infer policy goals in virtual market interventions, and whether policy changes achieved these goals.

\textit{Limitations}. 
Our study estimates causal effects of two main policy interventions on economic and trading activity in the OSRS virtual economy.
We conduct our analysis with data on item trading, over a number of months around new market interventions.
As such, we are able to estimate causal effects on trading prices and volume at the item-level, in the short to medium time-run.
As we do not have data on player-level interactions, we cannot further analyze the effects on player behavior as a whole.
Furthermore, we only observe prices in the real-world trading market (for virtual currency), and cannot analyze the prevalence of real-world trading directly.
Lastly, our data cover the majority of OSRS trading data, though not the whole market, as it is sourced in partnership with RuneLite.

While we do not use player-level data, our analysis plausibly identifies causal estimates for the effect of market interventions in the OSRS economy.
Uniquely, we did so for interventions that were implemented in the past, without our input on their design.
Thus, we performed no researcher-guided randomization or A/B design in our causal study.
Instead, we leveraged context-specific information regarding the design of the market intervention, combined with methodology common in empirical economics.
Firms that operate virtual (or even real-world) marketplaces routinely partner with researchers to study the effects of the prospective policy change, but these arrangements are not without costs.
Researchers must rely on firms to access their data and marketplace, researcher-guided randomization of new policies are often very costly to implement, and researchers must clear ethical guidelines when giving a new policy change to customers operating in such marketplaces.
While our approach has a number of limitations, our approach has the benefit of avoiding these costs, while maintaining a causal interpretation.
Thus, we did not deal with such problems or costs common for similar, causal analyses of market interventions.
To our knowledge, this is the first analysis to apply such methods to the setting of a virtual economy;
we set the stage for further work to utilize these methods for the study of virtual economies, overcoming problems related to traditional researcher-guided randomization and A/B testing.

\textit{Future Work}.
There are several avenues of future work as we believe our approach is generalizable across many games, since item exchanges are common in many online games. 
For example, World of Warcraft, a popular MMORPG, maintains an ``auction house'' that serves much of the same functions that the Grand Exchange offers. 
These online exchanges also may institute different policies. 
For example, in Guild War 2's auction house, there exists a fee to list an item for sale (5\% of sale price), and the exchange takes a 10\% portion of every sale. 
In both of the former games, the exchanges are global to the server. 
EVE Online, a space-themed MMORPG, maintains distributed marketplaces in the various ``star systems''. 
The EVE Online market also has various fees, such as a broker's fee, a sales fee or a relisting fee. 
Changes in any of the aforementioned fees may serve as unique natural experiments to perform studies like ours. 
To acquire the data necessary for these macro-level studies, one may consult official game APIs or those run by third-party community operators. 

Our approach need not only be applied to MMORPG economies, but may also be generally applied to online marketplaces. 
Many online marketplaces routinely update their platforms to institute different fees. 
For example, consider eBay, an online marketplace which charges insertion fees (for listing creation) and final value fees (for an item sale). 
These fees vary by item and by final sale value.
On October 10, 2022, eBay will change their sneakers sales fees, specifically for sneakers sold between \$100 and \$150~\cite{ebay-fees}. 
Accordingly, like in our analysis, there exist many discontinuities in the way a marketplace like eBay structures its fees that may be worthy of analysis. 
Another type of marketplace to consider are cryptocurrency markets, which also demonstrate rapidly changing fees and rules due to dynamic regulatory environments~\cite{shanaev2020taming}. 
Given the increasing levels of ecommerce and cryptocurrency activity, it will be important to understand how such markets react according to changing policies. 
Increasingly so, regulatory agencies, such as the Security and Exchange Commission have looked to MMORPG economies as units of study~\cite{down_2021}.
Virtual worlds have also been noted for their potential to study other concepts such as in epidemiology~\cite{lancetvirtual} or social networks~\cite{sciencevirtual}.


\section{Conclusion}
\label{sec:conclusion}
This work analyzes the impacts of various market interventions in a large virtual economy.
Specifically, this study concerns a transaction tax and an item sink introduced into the main item exchange in OSRS, a popular MMORPG.
Utilizing quasi-experimental techniques commonly used in empirical economics, we find that while the market interventions did lead to inflation in the price of the targeted items via the item sink, the trading volume of these items was not meaningfully affected. 
Furthermore, we also note that the real-world trading market remained relatively unaffected by the market interventions.
Finally, we also discuss the parallels between real-world economic phenomena and those witnessed in large-scale game economy such as OSRS. 
Our approach reinforces the idea that virtual economies and marketplaces may also be a useful tool in analyzing real-world economic theory.
Since game designers and marketplace operators routinely introduce policy updates, we see much potential for future studies, like ours, that study economic phenomena through large virtual economies and quasi-experimental methods.

\begin{acks}
This collaboration has been supported by a grant from Capital One. 
Silva’s research has also been supported by NASA; NSF awards CNS-1229185, CCF1533564, CNS-1544753, CNS-1730396, CNS-1828576, CNS-1626098; and DARPA PTG and D3M.
Any opinions, findings, and conclusions or recommendations expressed in this material are those of the authors and do not necessarily reflect the views of DARPA, NSF, NASA, or Capital One.
The authors would like to thank the maintainers of the OSRS Wiki for maintaining GE API endpoints and documentation, as well as providing crucial guidance.
\end{acks}

\bibliographystyle{ACM-Reference-Format}
\bibliography{00_references}

\clearpage
\appendix
\setcounter{table}{0}
\renewcommand{\thetable}{A\arabic{table}}
\setcounter{figure}{0}
\renewcommand{\thefigure}{A\arabic{figure}}
\section{Appendix}

\subsection{Regression Discontinuity}
\label{sec:rd}

The Regression Discontinuity (RD) design~\cite{hahn2001identification} is a method devised in econometrics to identify the causal effect of a treatment (often a change in policy), exploiting the fact that some policies are provided discontinuously according to a running variable.
The crucial assumption is that outcomes are continuously identical around the cut-off point, and their only difference when converging to the cut-off point above and below is treatment status.
In our case, we use the RD approach to analyse the effect of the new transaction tax on (log) trading volume.
Specifically, the tax is 1\% if price is above 100 GP (and below 500 million GP), and 0\% if below.

We estimate equation~\eqref{eqn:rd}, where $P_{i,t}$ refers to item $i$'s price on day $t$, $\text{Volume}_{i,t}$ its trading volume, $\mathbf{1} \left\{ . \right\}$ the indicator function, and $\varepsilon_{i,t}$ the error term.
$\text{Volume}_{i,t}$ counts the quantity of item $i$ traded on day $i$, and price $P_{i,t}$ is measured in virtual currency (GP).
$f(.)$ represents an unknown nuisance function of how volume relates to item price, which is estimated separately above and below the cut-off point.
\begin{equation}
    \label{eqn:rd}
    \log \left( \text{Volume}_{i,t} \right) =
        \alpha + f \left(P_{i,t} \right) +
        \beta \mathbf{1} \left\{ P_{i,t} \geq 100 \right\} + \varepsilon_{i,t}
\end{equation}
Parameter $\beta$ is the estimand, in this case the local average treatment effect (at $P_{i,t} = 100$) of the 1\% transaction tax on trading volume.
In practice, we estimate a local polynomial above and below the cut-off point, and the discontinuity between the two.
We restrict our data to observations within 20 GP of the cut-off (i.e. 80 to 120 GP), and for the month following the policy announcement, dates 09 December 2021 to 01 January 2022.
We implement this procedure via \textit{rdrobust} \cite{rdrobust,calonico2015optimal} for \textit{R} \cite{R}.

\begin{figure}[H]
    \includegraphics[width=\linewidth]{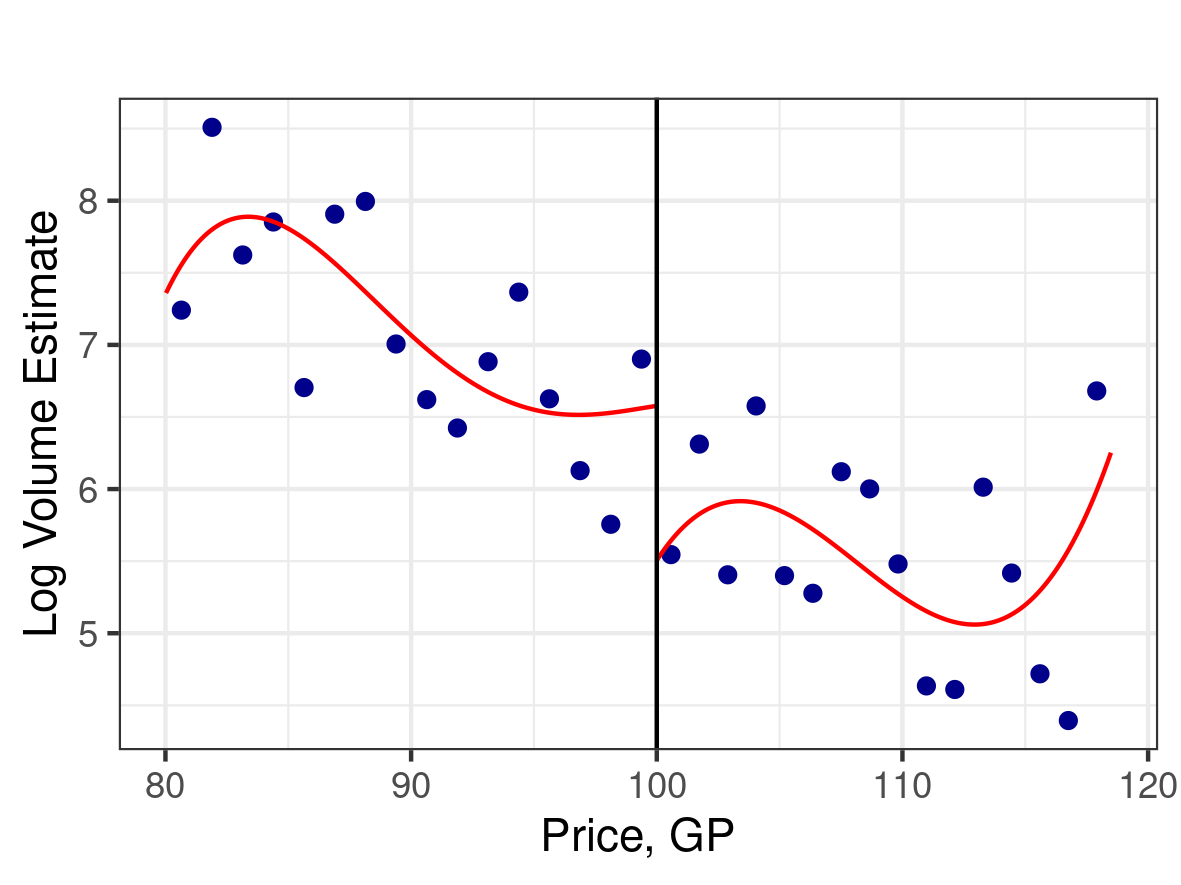}
    \caption{RD Results, Sharp around Low-End Tax Cut-off.}
    \label{fig:after-tax}
\end{figure}

The estimand is the distance between the two red lines in \autoref{fig:after-tax}; the red lines are the estimated local polynomial above and below the cut-off, and the dots are average trading volume for (binned) prices (and have associated weights which are not in the figure).
In this case, the distance between the lines at $P_{i,t} = 100$ is not statistically distinguishable from zero, and so we accept the null hypothesis that the transaction tax did not affect items' trading volume (around price of 100 GP). 

\subsection{Regression Kink}
\label{sec:rk}

The Regression Kink (RD) design~\cite{card2017regression} is a method devised in econometrics to identify the causal effect of a treatment (often a change in policy), exploiting the fact that some policies (mostly taxes) are provided at different rates (discontinuously) above and below according to a correlated variable.
The crucial assumption is that outcomes are continuously identical around the cut-off point, and their only difference when converging to the cut-off point above and below is treatment rate.
In our case, we use the RK approach to analyse the effect of the new transaction tax on (log) trading volume.
Specifically, the tax is 1\% if price is below 500 million GP (above 100 GP), and 5 million GP if above.
This variation leads to a change in rate of tax above and below the 500 million GP cut-off (see \autoref{fig:taxrate}).

We estimate equation~\eqref{eqn:rk}, where $P_{i,t}$ refers to item $i$'s price on day $t$, $\text{Volume}_{i,t}$ its trading volume, $T_{i,t}$ the new transaction tax, and $\varepsilon_{i,t}$ the error term.
$g(.)$ represents an unknown nuisance function of how volume relates to item price, which is estimated separately above and below the cut-off point.
\begin{equation}
    \label{eqn:rk}
        \begin{split}
        \log \left( \text{Volume}_{i,t} \right) =&
            \gamma + g \left(P_{i,t} \right) + \delta T_{i,t} + \varepsilon_{i,t} \\
        \text{where } T_{i,t} =&
            \mathbf{1} \left\{ P_{i,t} < 500 \times 10^6 \right\} \frac{P_{i,t}}{100} \\
            &+ \mathbf{1} \left\{ P_{i,t} \geq 500 \times 10^6 \right\} \left( 5 \times 10^6 \right)
    \end{split}
\end{equation}

Parameter $\delta$ is the estimand, in this case the local average treatment effect (at $P_{i,t} = 500 \times 10^6$) of the 1\% transaction tax on trading volume.
In practice, we estimate a local polynomial above and below the cut-off point, and the discontinuous difference between their slopes.
We restrict our data to observations with price greater than 100 million GP, and for the month following the policy announcement, dates 09 December 2021 to 01 January 2022.
We implement this procedure via \textit{rdrobust} \cite{rdrobust,calonico2015optimal} for \textit{R} \cite{R}.

\begin{figure}[H]
    \includegraphics[width=\linewidth]{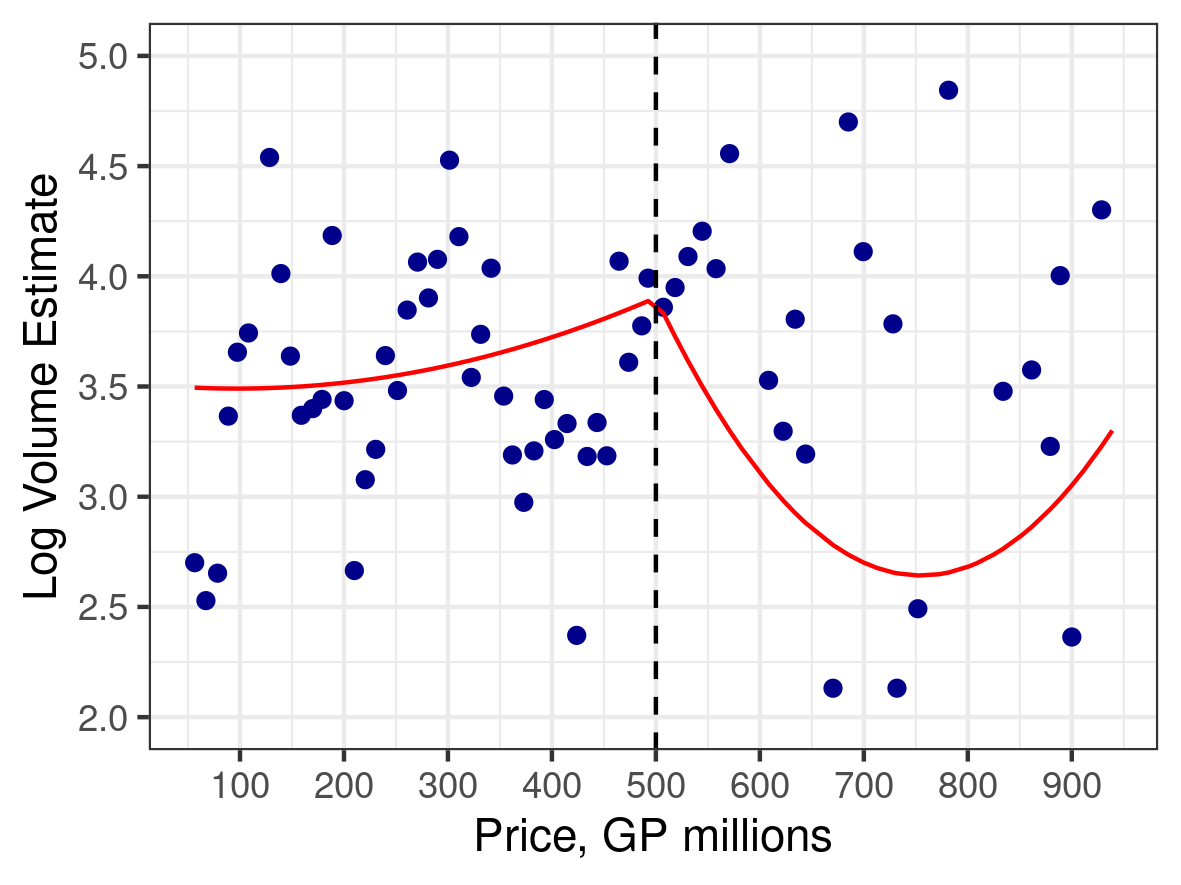}
    \caption{RK Results, around High-End Tax Cut-off.}
    \label{fig:kink-tax}
\end{figure}

The estimand is the slope differences between the two red lines in \autoref{fig:kink-tax} at the cut-off point; the red lines are the estimated local polynomial above and below the cut-off, and the dots are average trading volume for (binned) prices (and have associated weights which are not in the figure).
In this case, the difference in slope between the lines at $P_{i,t} = 500 \times 10^6$ is statistically distinguishable from zero, and so we reject the null hypothesis that the transaction tax did not affect items' trading volume (around price of 500 million GP).

\subsection{Difference in Difference}
\label{sec:did}

The Difference in Difference (DiD) design~\cite{card1994,bertrand2004much} is a method devised in econometrics to identify the causal effect of a treatment, when there is a plausible control group who did not receive the treatment.
The crucial assumption is that outcomes are the same for treatment and control groups follow parallel trends before the treatment was provided.
In our case, we use the DiD approach to analyse the effect of the new item sink on items' trading prices and trading volume.

\begin{figure}[H]
    \includegraphics[width=\linewidth]{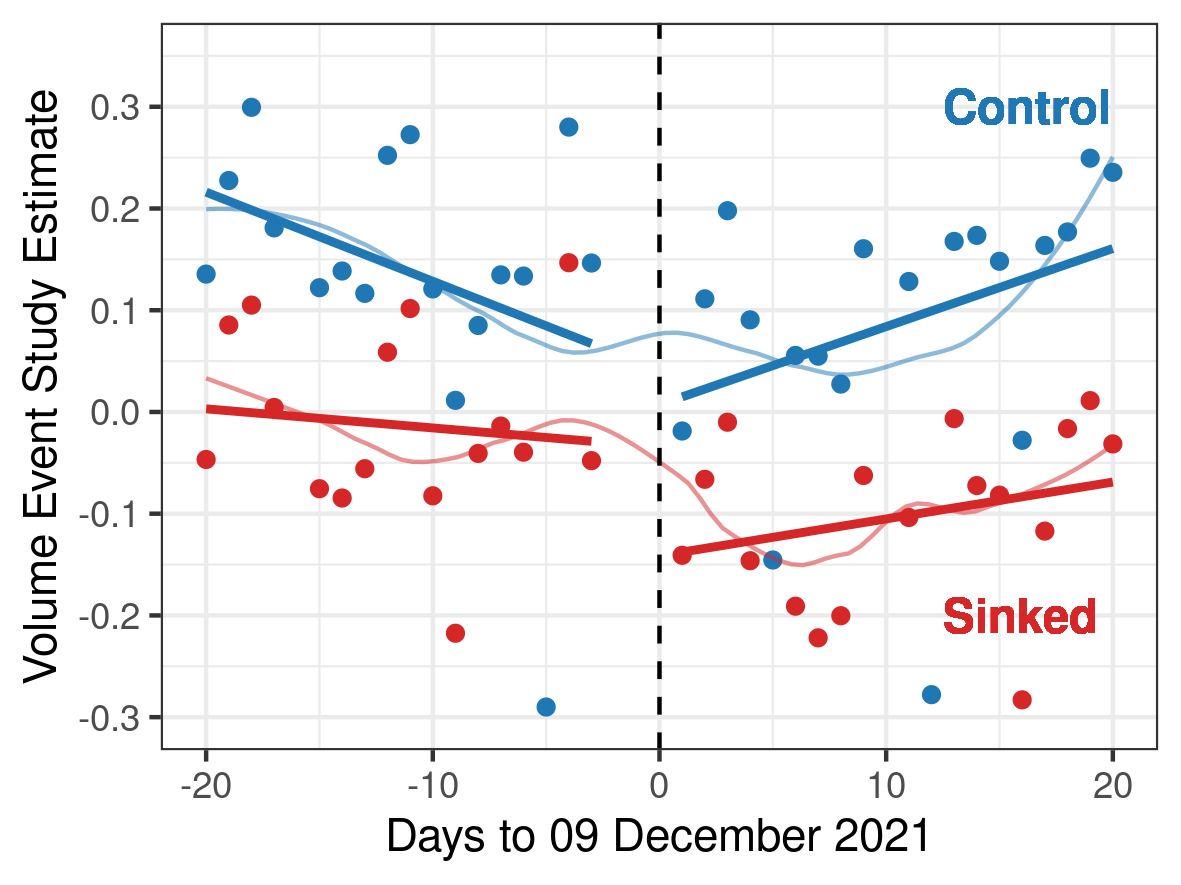}
    \caption{Pre-Trends in Volume, around New Item Sink.}
    \label{fig:volume-pretrends-sink}
\end{figure}

We show that both the treatment and control group have similar pre-trends before the item-sink was implemented in \autoref{fig:price-pretrends-sink}, \ref{fig:volume-pretrends-sink}.

For the RD model, we estimate \autoref{eqn:did}, where outcome $Y_{i,t}$ is an item's trading price or trading volume.
$\text{Post}_{i,t}$ is a binary for whether the observation is after the policy implementation date, 09 December 2021, $\text{Treated}_{i,t}$ a binary for whether the item is targeted by the new item sink.
Notably, we include fixed effects at the item-level, $\zeta_i$, in order to average out systematic item differences in price and trading volume --- and this term identifies treatment status, so that a binary for $\text{Treated}_{i,t}$ cannot be including separately.

\begin{equation}
    \label{eqn:did}
    \log \left( Y_{i,t} \right) =
        \zeta_i + \phi \text{Post}_{i,t} +
            \theta_{\text{DiD}} \text{Post}_{i,t} \times \text{Treated}_{i,t} + \varepsilon_{i,t}
\end{equation}

Parameter $\theta_{\text{DiD}}$ is the estimand, in this case the average treatment effect on the treated of the item sink on price and trading volume.
In practice, we estimate via linear regression, restricting to observations which either our in our constructed control group~\eqref{eqn:control} or the item-sink targeted (separately for our tax).
We only consider observations for after the policy announcement, 09 November 2019, and the month following the policy implementation, until 01 January 2022.

\begin{table}[h!]
    \caption{DiD Effects on Price and Volume for Initial and Secondary Sink.}
    
\begin{tabular}{@{\extracolsep{5pt}}lcccc} 
\\[-1.8ex]\hline 
\hline \\[-1.8ex] 
 & \multicolumn{4}{c}{Dependent Variable:} \\ 
\cline{2-5} 
\\[-1.8ex] & Price & Volume & Price & Volume \\ 
\\[-1.8ex] & (1) & (2) & (3) & (4)\\ 
\hline \\[-1.8ex] 
Post- & 0.053 & 0.158 & 0.074 & 0.038 \\ 
  & (0.005) & (0.024) & (0.005) & (0.025) \\ 
Post- $\times$ Treated & 0.070 & $-$0.034 & 0.147 & $-$0.045 \\ 
  & (0.011) & (0.056) & (0.024) & (0.112) \\ 
 \hline \\[-1.8ex]
Outcome Mean: & 13.4 & 756.4 & 9.9 & 767 \\ 
Initial sink? & Initial & Initial & Second & Second \\ 
Item-level FEs? & Yes & Yes & Yes & Yes \\ 
Observations & 7,438 & 7,545 & 6,563 & 6,650 \\ 
\hline 
\hline \\[-1.8ex] 
\end{tabular} 

    \label{tab:sink-did}
\end{table}

\begin{figure}[h!]
    \centering
    \includegraphics[width=\linewidth]{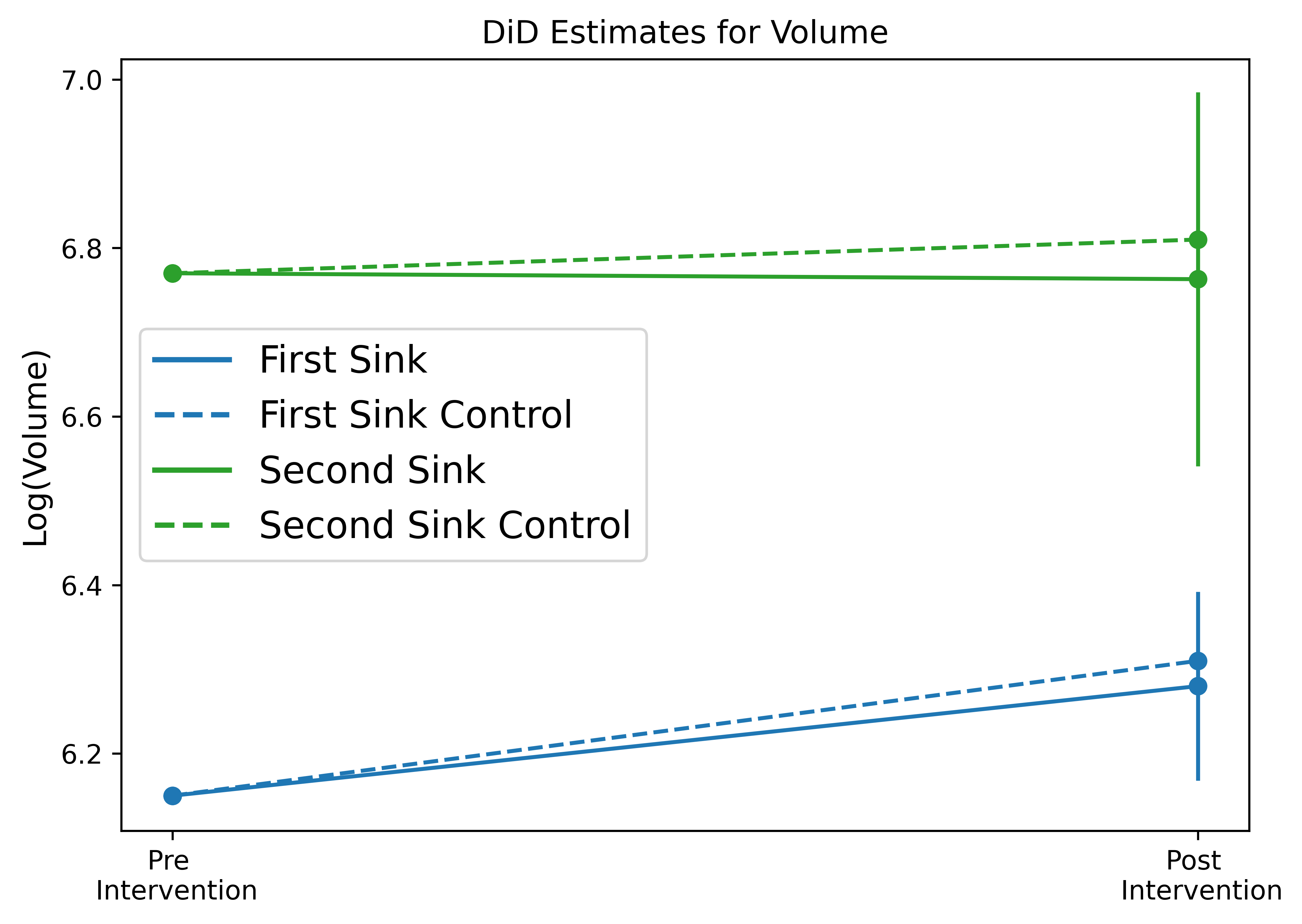}
    \caption{Lines are colored by \firstSink and \secondSink. Dotted lines indicate the counterfactual trend for market intervention, and vertical bars denote the confidence interval for the estimated effect. There was no detected effect on item trade volumes for sinked items.}
    \label{fig:did-volume}
\end{figure}

Our point estimate for the causal effect of the sink is presented in \autoref{tab:sink-did}, on the row labelled ``Post $\times$ Treated.''
The figures in brackets below each line are the associated standard errors of the point estimates, which are used to construct the point estimates in \autoref{fig:did} and \autoref{fig:did-volume}.
In this case, the sink is estimated to have positively affected price of the items in the sink (in both rounds of implementation), but not their trading volume.

\end{document}